\documentclass[prl,aps,twocolumn,amsfonts,showpacs,superscriptaddress]{revtex4-1}
\usepackage{epsfig} 
\usepackage{psfrag} 
\usepackage{amsmath}
\usepackage{amssymb} 
\usepackage{color} 
\usepackage{graphicx}
\usepackage{subfigure}

\begin{document}

\title{Ising Deconfinement Transition Between Feshbach-Resonant
  Superfluids} \author{S. Ejima} \affiliation{Institut f{\"u}r Physik,
  Ernst-Moritz-Arndt-Universit{\"a}t Greifswald, 17489 Greifswald,
  Germany.}  \author{M. J. Bhaseen} \affiliation{University of
  Cambridge, Cavendish Laboratory, Cambridge, CB3 0HE, UK.}
\author{M. Hohenadler} \affiliation{Institut f\"ur Theoretische Physik
  und Astrophysik, Universit\"at W\"urzburg, Germany.}
\author{F. H. L. Essler} \affiliation{The Rudolf Peierls Centre for
  Theoretical Physics, University of Oxford, Oxford, OX1 3NP.}
\author{H. Fehske} \affiliation{Institut f{\"u}r Physik,
  Ernst-Moritz-Arndt Universit{\"a}t Greifswald, 17489 Greifswald,
  Germany.}  \author{B. D. Simons} \affiliation{University of
  Cambridge, Cavendish Laboratory, Cambridge, CB3 0HE, UK.}

\begin{abstract}
  We investigate the phase diagram of bosons interacting via
  Feshbach-resonant pairing interactions in a one-dimensional
  lattice. Using large scale density matrix renormalization group and
  field theory techniques we explore the atomic and molecular
  correlations in this low-dimensional setting. We provide compelling
  evidence for an Ising deconfinement transition occurring between
  distinct superfluids and extract the Ising order parameter and
  correlation length of this unusual superfluid transition.  This is
  supported by results for the entanglement entropy which reveal both
  the location of the transition and critical Ising degrees of freedom
  on the phase boundary.
\end{abstract}

\date{\today}


\maketitle

The ability to cool atoms to low temperatures, and control their
interactions, has revolutionized the study of quantum many body
systems. Important achievements include realizations of Bose--Einstein
condensation (BEC), Bardeen--Cooper--Schrieffer (BCS) pairing in Fermi
gases, and strongly correlated Mott insulators (MIs).  In this development,
the BEC--BCS crossover between a gas of tightly bound molecules and
weakly bound Cooper pairs has played an instrumental role, and it has
been widely explored using Feshbach resonances to induce pairing. This
has led to diverse studies of the condensate fraction, single particle
gap, collective excitations, and vortices, and to pioneering
approaches to molecular quantum chemistry. For a review see
Ref.~\cite{Ketterle:Making}.

In recent work
\cite{Rad:Atmol,*Romans:QPT,*Radzi:Resonant,Lee:Bosefeshbach,*Lee:1DBF,Dickerscheid:Feshbach,*Sengupta:Feshbach,Gurarie:1DBFBR,*Gurarie:1DBF,Rousseau:Fesh,*Rousseau:Mixtures}
it has been argued that the BEC--BCS ``crossover'' for bosons is
strikingly different to the fermionic case since the atoms as well as
molecules may undergo Bose--Einstein condensation. These studies have
raised the exciting possibility of an Ising quantum phase transition
between distinct molecular (MC) and atomic plus molecular (AC+MC)
condensates.  In addition to discrete ${\mathbb Z}_2$ symmetry
breaking, this transition has a topological character and may be
viewed as a confinement-deconfinement transition for vortices.

The principal aim of this manuscript is to establish the presence of
such novel ${\mathbb Z}_2$ transitions in one-dimensional (1D) bosonic
Feshbach systems, where strong quantum fluctuations destabilize long
range superfluid order.  We combine large scale density matrix
renormalization group (DMRG) \cite{White:DMRG} and field theory
techniques to provide compelling evidence for Ising behavior. We
elucidate a full characterization of the scaling regime and the
proximate phases.  Our results demonstrate that an Ising transition
survives at strong coupling and large densities where field theory
arguments are no longer justified.  For related transitions in the
attractive Bose--Hubbard model with three--body losses see
Refs.~\cite{Daley:Three,*Daley:Threeerratum,*Diehl:Observability,Diehl:QFTI,*Diehl:QFTII},
and for analogues involving multicomponent fermions see
\cite{Wu:Competing,*Lecheminant:Confinement,Capponi:Confinement,*Roux:Spin}.
\begin{figure}[!h]
  \includegraphics[clip=true,width=3.2in]{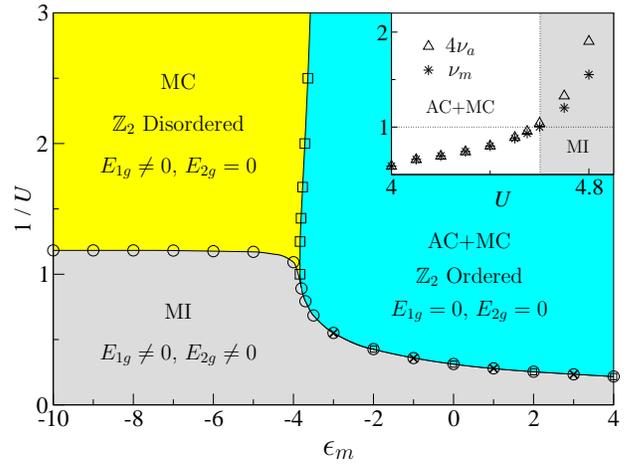}
  \caption{Phase diagram of the 1D Hamiltonian (\ref{atmolham}) with
    total density $\rho_{\rm T}=N_{\rm T}/L=2$, showing a Mott
    insulator (MI), a molecular condensate (MC), and a coupled atomic
    plus molecular condensate (AC+MC). We use DMRG with up to $L=128$
    and open boundaries, with $\epsilon_a=0$,
    $U_{aa}/2=U_{mm}/2=U_{am}=g=U$, $t_a=1$, $t_m=1/2$.  The squares
    and circles indicate the vanishing of the one-particle and
    two-particle gaps, $E_{1g}$ and $E_{2g}$, as
    $L\rightarrow\infty$. The crosses show where the molecular 
exponent, $\nu_m$, reaches unity.  Inset: AC+MC to MI
    transition at $\epsilon_m=4$. The atomic and molecular exponents,
    $\nu_a$ and $\nu_m$, are locked up to the MI boundary where
    $\nu_m=1$, indicating the absence of an AC phase.}
\label{Fig:PD}
\end{figure}

We consider the Hamiltonian
\cite{Dickerscheid:Feshbach,*Sengupta:Feshbach,Rousseau:Fesh,*Rousseau:Mixtures,Bhaseen:Feshising}
\begin{equation}
  \begin{aligned}
    H & =\sum_{i\alpha}\epsilon_\alpha n_{i\alpha}- \sum_{\langle
      ij\rangle}\sum_\alpha t_\alpha
    \left(b_{i\alpha}^\dagger b_{j\alpha}+{\rm H.c.} \right)\\
    & \hspace{1cm} +\sum_{i\alpha\alpha^\prime}
    \frac{U_{\alpha\alpha^\prime}}{2}:n_{i\alpha}n_{i\alpha^\prime}: +
    \,H_{{\rm F}},
\label{atmolham}
\end{aligned}
\end{equation}
describing bosons, $b_{i\alpha}$, hopping on a lattice with sites $i$,
where $\alpha=a,m$ labels atoms and molecules. Here, $\epsilon_\alpha$
are on-site potentials, $t_\alpha$ are hopping parameters, $\langle
ij\rangle$ denotes summation over nearest neighbor bonds, and
$U_{\alpha\alpha^\prime}$ are interactions.  Normal ordering yields
$:n_{i\alpha}n_{i\alpha}: = n_{i\alpha}(n_{i\alpha}-1)$ for like
species, and
$:n_{i\alpha}n_{i\alpha^\prime}:=n_{i\alpha}n_{i\alpha^\prime}$ for
distinct species. Molecules are formed by the Feshbach term, $H_{{\rm
    F}}= g\sum_i(m^\dagger_{i} a_{i}a_{i}+{\rm H.c.})$, where
$m_i\equiv b_{im}$ and $a_i\equiv b_{ia}$. Atoms and molecules are not
conserved, but the total, $N_{\rm T}\equiv \sum_{i}(n_{ia}+2n_{im})$,
is preserved.

To orient the discussion, we present a section of the phase diagram in
Fig.~\ref{Fig:PD}, with parameters chosen for comparison with previous
studies \cite{Rousseau:Fesh}.  In this manuscript we use DMRG on a 1D
system with up to $L=512$ sites, and adopt units where $t_a=1$.  We
allow up to five atoms and five molecules per site, and retain up to
$m_\rho=2400$ states in the density matrix so that the discarded
weight is less than $1\times 10^{-10}$. The phase boundaries
correspond to the vanishing of the one-particle and two-particle
excitation gaps, $E_{1g}\equiv \mu_{1p}(L)-\mu_{1h}(L)$ and
$E_{2g}\equiv \mu_{2p}(L)-\mu_{2h}(L)$ respectively, where
$\mu_{np}(L)=E_0(L,N_{\rm T}+n)-E_0(L,N_{\rm T})$,
$\mu_{nh}(L)=E_0(L,N_{\rm T})-E_0(L,N_{\rm T}-n)$, and $E_0$ is the
ground state energy.  The diagram shows a MI with
gaps for both excitations $E_{1g}\neq 0$ and $E_{2g}\neq 0$, a
MC phase with a one-particle gap $E_{1g}\neq 0$ and
$E_{2g}=0$, and a coupled atomic plus molecular condensate (AC+MC)
with $E_{1g}=0$ and $E_{2g}=0$. In contrast to the qualitative diagram
in Ref.~\cite{Rousseau:Fesh}, inferred from quantum Monte Carlo
simulations on smaller systems, we find no evidence for a
single-component AC phase. This is in accord with
expectations in higher dimensions
\cite{Rad:Atmol,Romans:QPT,Radzi:Resonant}.  As we will discuss, this
is supported by direct evaluation of correlation functions using both
DMRG and field theory.  Throughout the AC+MC phase we find power laws
for atoms {\em and} molecules with related exponents; see inset of
Fig.~\ref{Fig:PD}.  The conclusions of Ref.~\cite{Rousseau:Fesh} are
hampered by the slow divergence of the associated zero momentum
molecular occupation number with increasing $L$, close to the MI
boundary. This also afflicts the molecular visibility.  Here, our
focus is on the transition between the MC and AC+MC superfluids. We
begin with symmetry arguments and field theory predictions before
comparison with DMRG.

An intuitive way to understand the origin of the proposed Ising
transition between the MC and AC+MC phases is via the symmetry of the
Hamiltonian (\ref{atmolham}) under ${\rm U}(1)\times {\mathbb Z}_2$
transformations. This corresponds to invariance under $m\rightarrow
e^{i\theta}m$ and $a\rightarrow e^{i(\theta/2\pm \pi)}a$, where
$\theta\in {\mathbb R}$.  In general these symmetries may be broken
independently. Before discussing the problem in 1D, where continuous
${\rm U}(1)$ symmetry breaking is absent, let us first recall the
situation in higher dimensions
\cite{Rad:Atmol,*Romans:QPT,*Radzi:Resonant}.  In this case, the
molecular condensate (MC) phase has $\langle m\rangle\neq 0$ and
$\langle a\rangle=0$. This only breaks the ${\rm U}(1)$ contribution
and leaves the ${\mathbb Z}_2$ symmetry, $a\rightarrow -a$, intact;
this corresponds to the disordered phase of an Ising model, coexisting
with molecular superfluidity.  On the other hand, the coupled atomic
plus molecular condensate (AC+MC) phase has $\langle m\rangle \neq 0$
{\em and} $\langle a\rangle\neq 0$. This breaks the ${\rm U}(1)\times
{\mathbb Z}_2$ symmetry completely and corresponds to the ordered
phase of an Ising model, coexisting with atomic and molecular
superfluidity. Returning to the present 1D problem, where continuous
${\rm U}(1)$ symmetry breaking is absent, the spontaneous formation of
expectation values $\langle a\rangle$ and $\langle m\rangle$ is
prohibited. Instead, superfluid order is characterized by long range
power law correlations, and the nature of the phases and transitions
in Fig.~\ref{Fig:PD} requires closer inspection.

Owing to the ${\rm U}(1)\times {\mathbb Z}_2$ symmetry of the
Hamiltonian, the low energy Lagrangian of the MC to AC+MC transition
is given by ${\mathcal L}={\mathcal L}_\vartheta+{\mathcal
  L}_{\phi}+{\mathcal L}_{\vartheta\phi}$
\cite{Lee:Bosefeshbach,Radzi:Resonant}, where
\begin{equation} {\mathcal L}_\vartheta=
  \frac{K_\vartheta}{2}\left[c_\vartheta^{-2}(\partial_\tau\vartheta)^2
    +(\partial_x\vartheta)^2\right],
\label{scalar}
\end{equation}
is a ${\rm U}(1)$ invariant free scalar field, and 
\begin{equation} {\mathcal
    L}_\phi=\frac{K_\phi}{2}\left[c_\phi^{-2}(\partial_\tau\phi)^2
    +(\partial_x\phi)^2\right]-\eta\phi^2+\lambda\phi^4,
\end{equation}
is an Ising model in the soft-spin $\phi^4$ representation.  The
coupling, ${\mathcal
  L}_{\vartheta\phi}=i\phi^2\partial_\tau\vartheta/2$, has a similar
form to a Berry phase \cite{Lee:Bosefeshbach,Radzi:Resonant}.  A
similar action also emerges for quantum wires \cite{Sitte:Emergent}.
In the following we neglect ${\mathcal L}_{\vartheta\phi}$ and examine
the reduced theory.  Within mean field theory, ${\mathcal
  L}_{\vartheta\phi}\sim
i\langle\phi\rangle^2\partial_\tau\vartheta/2$ acts like a boundary
term, and this is expected to provide a good description of the
proximate phases. Near the transition, this cannot be neglected {\em a
  priori}, and ${\mathcal L}_{\vartheta\phi}$ may change the behavior
on very large length scales and in other regions of the phase diagram
\cite{Sitte:Emergent}. Nonetheless, we find excellent agreement with
bulk properties. The parameters $K_\vartheta$, $c_\vartheta$,
$K_\phi$, $c_\phi$, $\eta$, $\lambda$, are related to the coefficients
of $H$.  Atoms and molecules are described by the semiclassical
number-phase relations, $m\sim \sqrt{\rho_m}\,e^{i\vartheta}$, and
$a\sim \phi\,e^{i\vartheta/2}$, where $\rho_m$ is the molecular
density.  We will explore the consequences of this correspondence in
1D, for local observables and correlations.

Let us first gather consequences of this correspondence for local
observables. Deep within the ${\mathbb Z}_2$ disordered MC phase,
$\eta\gg 0$ and $\langle \phi(x)\rangle=0$.  However, $\phi^2(x)$ may
have a non-zero average.  It follows that the densities of atoms and
molecules, $\langle a^\dagger(x)a(x)\rangle\sim\langle
\phi^2(x)\rangle$ and $\langle m^\dagger(x)m(x)\rangle\sim \rho_m$,
are generically non-zero in {\em both} the AC+MC {\em and} MC
phases. In addition, $\langle m^\dagger(x)a(x)a(x)\rangle\sim
\sqrt{\rho_m}\,\langle\,\phi^2(x)\rangle$, acquires true long range
order, even in this 1D setting; $H_{\rm F}$ locks the atomic and
molecular condensates as encoded in the number-phase
relations. However, this local average is naively insensitive to the
${\mathbb Z}_2$ transition due to invariance under $a\rightarrow
-a$. Insight is better gleaned from correlations.

It follows from the relation $m\sim \sqrt{\rho_m}\,e^{i\vartheta}$,
that the molecular correlation function $\langle
m^\dagger(x)m(0)\rangle\sim \rho_m\langle
e^{-i\vartheta(x)}e^{i\vartheta(0)}\rangle \sim x^{-\nu_m}$ decays
like a power law, where $\nu_m=1/2\pi K_\vartheta$ varies throughout
the phase diagram. In contrast, the behavior of the atomic correlation
function, $\langle a^\dagger(x) a(0)\rangle\sim\langle
\phi(x)\phi(0)\rangle\langle
e^{-i\vartheta(x)/2}e^{i\vartheta(0)/2}\rangle \sim \langle
\phi(x)\phi(0)\rangle x^{-\nu_m/4}$, depends on the Ising prefactor.
We consider the disordered and ordered phases in turn.

\begin{figure}
  \includegraphics[clip=true,width=3.2in]{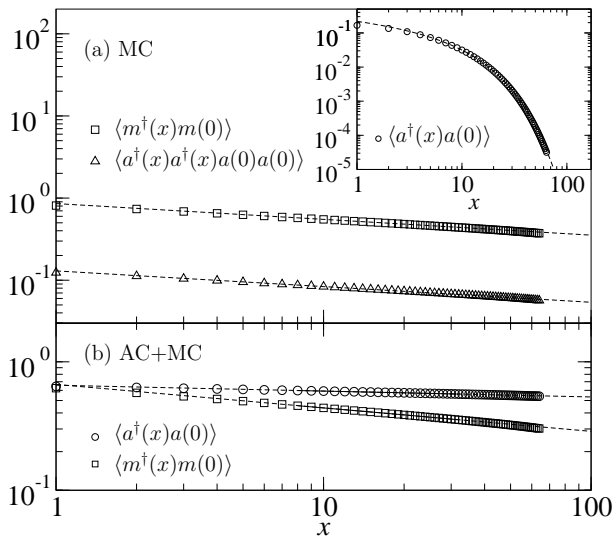}
  \caption{Correlation functions using DMRG with $L=128$ and
    open boundaries. We use the parameters in
    Fig.~\ref{Fig:PD} with $U=0.7$.  (a) ${\mathbb Z}_2$ disordered MC
    phase with $\epsilon_m=-4$, revealing power laws for molecules and
    atomic bilinears with the same exponent; the fits are
    $y=0.858\,x^{-0.1922}$ and $y=0.130\,x^{-0.1909}$.  Inset: Atomic
    correlations decay exponentially. We fit to the
    prediction $\langle a^\dagger(x)a(0)\rangle\sim x^{-\nu_m/4}{\rm
      K}_0(x/\xi)$, where we input $\nu_m$ from panel (a) and extract
    $\xi\approx 9.28$.  This establishes MC as a pairing phase without
    atomic condensation.  (b) ${\mathbb Z}_2$ ordered AC+MC phase with
    $\epsilon_m=-3$. Atoms and molecules exhibit power law exponents 
locked by a factor of four; the fits are $y=0.667\,
    x^{-0.1827}$ and $y=0.657\, x^{-0.0456}$. }
\label{Fig:Corr}
\end{figure}

In the ${\mathbb Z}_2$ disordered MC phase, the atomic correlation
function decays exponentially with a power law prefactor, $\langle
a^\dagger(x)a(x)\rangle \sim x^{-\nu_m/4}{\rm K}_0(x/\xi) \sim
x^{-1/2-\nu_m/4}e^{-x/\xi}$. Here we use the result for the hard-spin
Ising model, $\langle \phi(x)\phi(0)\rangle \sim {\rm K}_0(x/\xi)$,
where ${\rm K}_0$ is a modified Bessel function and $\xi$ is the Ising
correlation length \cite{Wu:Spin}.  On the other hand, {\em pairs} of
atoms condense and exhibit power law correlations, $\langle
a^\dagger(x)a^\dagger(x)a(0)a(0)\rangle \sim \langle \phi^2\rangle^2\,
x^{-\nu_m}$, with the {\em same} exponent as the molecular two-point
function, $\nu_m$. That is to say, the MC phase is a pairing phase of
bosons without single particle condensation \cite{Valatin:Collective}.
In order to test these weak coupling predictions we perform DMRG on
the 1D Hamiltonian (\ref{atmolham}).  As predicted, this behavior is
well supported by our simulations in Fig.~\ref{Fig:Corr}(a),
which reveal identical power laws for molecules and atomic bilinears,
with exponential decay for atoms.  This behavior extends throughout
the MC phase, including the Mott boundary in the strongly coupled
regime.

In contrast, in the ${\mathbb Z}_2$ ordered AC+MC phase, both
molecules {\em and} atoms have power law correlations, $\langle
m^\dagger(x)m(0)\rangle\sim x^{-\nu_m}$, $\langle
a^\dagger(x)a(0)\rangle\sim \langle\phi\rangle^2\,x^{-\nu_a}$, with
locked exponents, $\nu_m=4\nu_a$, that differ by a factor of four
\cite{Lee:Bosefeshbach,Radzi:Resonant}. Again, these features are
readily seen from our large scale DMRG simulations in
Fig.~\ref{Fig:Corr}(b) and the inset of Fig.~\ref{Fig:PD}. Likewise,
this behavior persists into the strong coupling limit, where the field
theory approach no longer strictly applies. In particular, we have
checked that the molecular correlation function, $\langle
m^\dagger(x)m(0)\rangle\sim x^{-\nu_m}$, remains a power law
throughout the AC+MC phase and close to the Mott boundary in
Fig.~\ref{Fig:PD}.  This is consistent with the absence of an AC phase
\cite{Rad:Atmol,Romans:QPT,Radzi:Resonant} in contrast to
Ref.~\cite{Rousseau:Fesh}. The latter employ the zero-momentum
occupation, $n(0)$.  However, the Fourier transform of $x^{-\nu}$
gives $n(0)\sim {\rm const} + {\rm const}\,L^{1-\nu}$; close to the
MI where $\nu_m=1$ one may miss the slow divergence of $n_m(0)$.

\begin{figure}
\includegraphics[clip=true,width=3.2in]{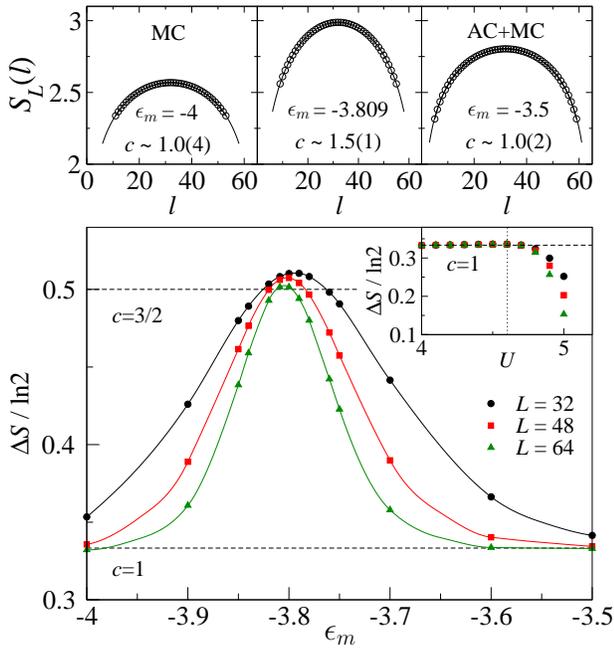}
\caption{Top: Entanglement entropy $S_L(l)$ obtained by DMRG in a
  periodic system with $L=64$.  We transit through the AC to AC+MC
  transition in Fig.~\ref{Fig:PD} with $U=0.7$.  The fits to
  Eq. (\ref{ee}) yield $c\approx 1$ in the MC and AC+MC phases, and
  $c\approx 3/2$ close to the transition. This reflects additional
  critical ${\mathbb Z}_2$ degrees of freedom. Due to the asymptotic
  nature of Eq.  (\ref{ee}), high quality fits are obtained from the
  central region away from the boundaries.  Bottom: Entanglement
  entropy difference $\Delta S(L)$ showing an Ising transition at
  $\epsilon_m\approx-3.8$ for $U=0.7$, in agreement with
  Fig.~\ref{Fig:PD}. The solid lines are spline fits. Inset: $\Delta
  S$ on passing through the AC+MC to MI transition at $\epsilon_m=4$,
  suggesting an XY transition with $c=1$.}
\label{Fig:EE}
\end{figure}

Having established a close connection between field theory and DMRG
for the MC and AC+MC phases, let us now examine the transition.  A key
diagnostic is the central charge, $c$, which counts critical degrees
of freedom. This may obtained from the entanglement entropy. For a
block of length $l$ in a periodic system of length $L$, the von
Neumann entropy is given by $S_L(l)=-{\rm Tr}_l(\rho_l\ln \rho_l)$,
where $\rho_l={\rm Tr}_{L-l}(\rho)$ is the reduced density matrix.
One obtains \cite{Holzey:Entropy,*Calabrese:Entanglement}
\begin{equation}
S_L(l)=\frac{c}{3}\ln\left[\frac{L}{\pi}
\sin\left(\frac{\pi l}{L}\right)\right]+s_1,
\label{ee}
\end{equation}
where $s_1$ is a constant. As may be seen in Fig.~\ref{Fig:EE}, the
numerically extracted central charge of the MC phase yields $c=1$, as
one would expect for a free boson, with coexisting gapped degrees of
freedom.
In addition, the AC+MC phase also has $c=1$. Note that it is {\em not}
$c=2$ as would be the case for two independent Luttinger liquids. This
reflects the coupled nature of the atomic and molecular condensates in
the AC+MC phase, with additional gapped Ising degrees of freedom; the
Feshbach term is relevant and drives the ${\mathbb Z}_2$ sector
massive. Close to the MC to AC+MC transition, where the anticipated
Ising gap closes, one expects the central charge to increase to
$c=3/2$, due to {\em additional} critical Ising degrees of freedom
with $c=1/2$.  This is confirmed by DMRG in Fig.~\ref{Fig:EE}. Further
evidence is obtained from the difference \cite{LK:Spreading}, $\Delta
S(L)\equiv S_L(L/2)-S_{L/2}(L/4)=\frac{c}{3}\ln(2)+\dots$, as a
function of $\epsilon_m$; see Fig.~\ref{Fig:EE}. For a given $L$ this
displays a peak, whose location coincides with the MC to AC+MC
transition obtained via the single-particle gap in
Fig.~\ref{Fig:PD}. The evolution with increasing $L$ is consistent
with the passage towards $c=1$ in the superfluid phases, and $c=3/2$
in the vicinity of the transition. Application of this method to the
MI to superfluid transitions
\cite{Dickerscheid:Feshbach,*Sengupta:Feshbach} yields $c=1$ close to
the MI boundary, suggesting XY behavior; see inset of
Fig.~\ref{Fig:EE}.  The absence of criticality within the MI phase is
evidence against a super--Mott state \cite{Rousseau:Fesh} and
correlations decay exponentially
\cite{Bhaseen:Feshising,Eckholt:Comment}.
 
\begin{figure}
  \includegraphics[clip=true,width=3.2in]{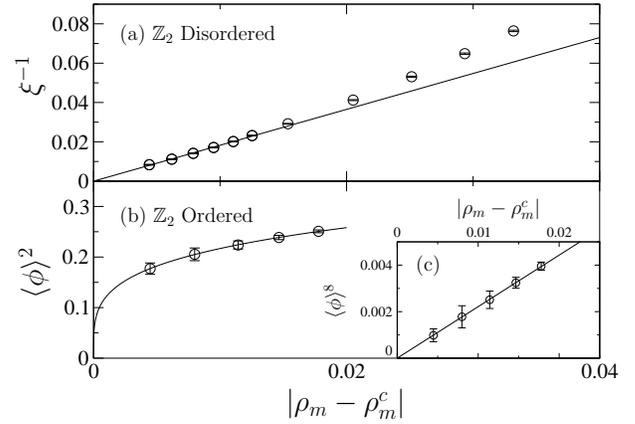}
  \caption{DMRG with open boundaries. We use the parameters in
    Fig.~\ref{Fig:PD} with $U=0.7$ and transit from MC to AC+MC.
    ${\mathbb Z}_2$ disordered MC phase: (a) $\xi$ extracted from the
    atomic correlations with $L=256$. $\xi^{-1}\sim
    |\rho_m-\rho_m^c|^{\nu}$, where $\rho_m$ ($\rho_m^c\approx 0.85$)
    is the (critical) density of molecules and $\nu=1$.  ${\mathbb
      Z}_2$ ordered AC+MC phase: (b) $\langle \phi\rangle^2$, up to a
    constant prefactor, obtained by finite size scaling of the atomic
    correlations with up to $L=512$. (c) Re-plotting yields
    $\beta=1/8$.}
\label{Fig:Isingcrit}
\end{figure}

Having provided evidence for a ${\mathbb Z}_2$ superfluid transition,
we now extract the Ising correlation length, $\xi$, and order
parameter, $\langle \phi\rangle$, via finite size scaling of the
atomic and molecular correlations.  Due to the absence of particle
conservation, and the presence of additional superfluid degrees of
freedom, these cannot be readily obtained from the energy spectra
alone.  Ising scaling close to the transition implies that
$\xi^{-1}\sim |{\mathcal M}-{\mathcal M}_c|$ and $\langle
\phi\rangle\sim |{\mathcal M}-{\mathcal M}_c|^{1/8}$, where ${\mathcal
  M}$ is a mass scale parametrizing the departure from
criticality. We identify the molecular density, ${\mathcal M}\sim
\rho_m$, as the appropriate scaling variable.  As shown in
Fig.~\ref{Fig:Isingcrit}, the DMRG results are in excellent agreement
with Ising critical exponents. This is non-trivial since the Ising
degrees of freedom are non-local with respect to the atoms and
molecules themselves.
In summary, we have studied bosons interacting via Feshbach
interactions in a 1D lattice. We provide evidence for an Ising quantum
phase transition between distinct superfluids. We extract both the
${\mathbb Z}_2$ order parameter, $\langle \phi\rangle$, and the Ising
correlation length, $\xi$.  It would be interesting to see if this
${\mathbb Z}_2$ transition may be driven first order, and the effect
of higher bands \cite{Buchler:Micro}.  One may also consider ${\mathbb
  Z}_{\rm N}$ transitions involving ${\rm N}$-particle pairing.

We thank F. Assaad, S. Capponi, N. Cooper, S. Diehl, M. Garst,
Z. Hadzibabic, E. Jeckelmann, M. K\"ohl, A. Lamacraft, A. L{\"a}uchli,
C. Lobo and A. Silver for discussions.  MJB and BDS acknowledge EPSRC
grant no.  EP/E018130/1. FHLE by EP/D050952/1.  SE and HF acknowledge
DFG grant SFB 652. MH by DFG FG1162.

\end{document}